\documentstyle[preprint,aps]{revtex}
\tightenlines

\def\R{\mbox{\boldmath $R$}}
\def\r{\mbox{\boldmath $r$}}
\def\P{\mbox{\boldmath $P$}}
\def\p{\mbox{\boldmath $p$}}
\def\q{\mbox{\boldmath $q$}}

\def\J{\mbox{\boldmath $J$}}
\def\A{\mbox{\boldmath $A$}}
\def\B{\mbox{\boldmath $B$}}
\def\tt{\mbox{\boldmath $\tau$}}
\def\ss{\mbox{\boldmath $\sigma$}}
\def\dd{\mbox{\boldmath $\nabla$}}

\begin{document}
\begin{center}
{\Large\bf{Photon-induced two-nucleon knockout reactions to discrete final 
states}}
\end{center}
\vskip0.2truecm
\centerline{\large{C. Giusti and F.D. Pacati}}
\vskip0.5truecm
\centerline{\small{\em {{Dipartimento di Fisica Nucleare e Teorica, 
Universit\`a di Pavia, and}}}}

\centerline{\small{\em {{Istituto Nazionale di Fisica Nucleare, Sezione di
Pavia, Pavia, Italy}}}}

\baselineskip=0.75cm
\vskip1.0truecm

\section*{}

\noindent
{\small {\bf{Abstract}}}\\

\baselineskip=0.5cm
{\small {Cross sections and photon asymmetries of the exclusive 
$^{16}\mathrm{O}(\gamma,\mathrm{pn})^{14}\mathrm{N}$ and 
$^{16}\mathrm{O}(\gamma,\mathrm{pp})^{14}\mathrm{C}$ knockout reactions are 
calculated for transitions to the low-lying discrete final states of the 
residual nucleus in the photon-energy range between 100 and 400 MeV. Exclusive 
reactions may represent a test of reaction mechanisms and a promising tool for 
investigating the dynamics of nucleon pairs in different states. Cross 
sections and asymmetries for both ($\gamma$,pn) and ($\gamma$,pp) turn out to 
be only slightly affected by short-range correlations and dominated by 
two-body currents. Therefore, two-nucleon knockout reactions induced by real
photons appear well suited to investigate the nuclear current and the 
selectivity of individual transitions to its different components.\\
\\
\noindent
PACS numbers: 25.20.Lj, 24.50+g}}

\noindent


\vspace{5mm}

\section*{1. Introduction}
\baselineskip=0.75cm

\vspace{-5mm}

\hspace*{6mm}

\tightenlines
Electromagnetically induced two-nucleon knockout reactions have long been 
considered an important tool for investigating the properties of nucleon 
pairs within nuclei and their interaction at short 
distance~[\ref{Gottfried},\ref{Oxford}]. 

Direct information on dynamical short-range correlations (SRC), which are 
linked to the short-ranged repulsive core of the NN--interaction, can be 
obtained if one assumes that the real or virtual photon hits, through a 
one-body current, either nucleon of a correlated pair and both nucleons are 
then ejected. However, nucleon pairs can also be ejected by two-nucleon 
currents, which effectively take into account the influence of subnuclear 
degrees of freedom like mesons and isobars. A reliable treatment of both these 
competing processes as well as of other reaction mechanisms is in principle 
needed before one can draw definite conclusions in comparison with data. 
However, since the role and relevance of these contributions is different in 
different reactions and kinematics, it is possible to envisage appropriate 
situations where various specific effects can be disentangled and separately 
investigated. 

The most promising tool for studying SRC is represented by the (e,e$'$pp) 
reaction, where the effect of two-body currents is less dominant as compared 
to the (e,e$'$pn) and ($\gamma$,NN) processes. The first exploratory triple 
coincidence measurements of the (e,e$'$pp) cross section, performed at NIKHEF 
on $^{12}{\mathrm {C}}$~[\ref{Kester},\ref{Zondervan}], gave clear evidence of 
two-nucleon knockout in the dip region. Peaks corresponding to knockout of 
protons from the $1p$ and $1s$ shells were clearly recognized in the 
missing-energy spectrum. The measured cross sections were found in 
satisfactory agreement with the model calculations of 
refs.~[\ref{GP},\ref{BGPR}] in the region corresponding to $(1p)^2$ knockout 
and indicated that a large part of the strength can be attributed to SRC. 
In most recent experiments on $^{16}{\mathrm {O}}$ at 
NIKHEF~[\ref{Gerco},\ref{Gercoth}] and Mainz~[\ref{Rosner}] it was possible to 
achieve sufficient energy resolution to allow the separation of the cross 
section related to distinct states of $^{14}{\mathrm {C}}$. A further 
experiment on $^{16}{\mathrm {O}}$ with improved statistics has been approved 
in Mainz~[\ref{MAMI}]. The presence of discrete final states with well-defined 
angular momentum makes $^{16}{\mathrm {O}}$ a particularly attractive target, 
since specific final states may act as filter for the study of different 
reaction processes involving SRC and two-nucleon 
currents~[\ref{eepp},\ref{sf}]. Moreover, the first and so far only available 
calculation of the two-nucleon spectral function for a finite nucleus, 
has been applied just to $^{16}{\mathrm{O}}$~[\ref{Geurts}]. The corresponding 
pair removal amplitudes have been included in the description of the 
(e,e$'$pp) knockout reaction~[\ref{sf}]. The predicted selectivity of the 
considered reaction involving different final states, which is confirmed by 
the fair agreement obtained in the first comparisons with 
data~[\ref{Gercoth},\ref{Rosner}], opens up good perspectives for the study of 
SRC  in (e,e$'$pp) reactions. 

The ($\gamma$,NN) reactions appear dominated by medium-range 
meson-exchange currents and a good understanding of the dominant 
single-pion-exchange mechanisms is essential before one can hope of 
investigating shorter range effects~[{\ref{Ryck},\ref{Ryck1}]. 

First exploratory ($\gamma$,NN) experiments performed in Bonn~[\ref{Bonn}] and 
Tokio~[\ref{Tokio}] had neither good statistical accuracy nor sufficient 
energy resolution to allow a straightforward investigation of direct processes.
Early work in Mainz~[\ref{MacG}] had good energy resolution allowing the 
identification of nucleons emitted from different shells. Further experiments 
with good statistics and resolution were carried out in
Mainz~[\ref{Mainz},\ref{Mainz98}] and Lund~[\ref{Lund}] in order to disentangle 
the reaction mechanisms. A major conclusion was obtained through comparison of 
missing-energy spectra with the results of the theoretical model of 
ref.~[\ref{Oset}], where different contributing mechanisms are included in the 
description of the pion and nucleon photoproduction channels in a unified 
diagrammatic approach: only the low missing-energy region 
($E_{\mathrm{2m}}\leq 40$ MeV) is predominantly fed by direct photoabsorption 
on a nucleon pair, whereas for $E_{\mathrm{2m}}\geq 70$ MeV other mechanisms, 
e.g. contributions originating from quasifree ($\gamma$,$\pi$) production and 
subsequent pion final-state interactions, become dominant. This result implies 
that only low values of $E_{\mathrm{2m}}$ must be considered when studying the 
properties of nucleon pairs and was qualitatively confirmed by the analysis of 
the data in ref.~[\ref{Mainz98}] through the model of ref.~[\ref{Ryck}].

Measurements with sufficient energy resolution to separate the low-lying final 
states of the residual nucleus are of great interest also in the case of 
photon-induced reactions. They represent a stringent test of reaction 
models and allow one to investigate the state dependence of the photoabsorption 
process on nucleon pairs. However, most of the data collected up till now 
could not separate individual final states and simultaneously provide 
dependences of the cross sections on dynamical variables. The 
high-resolution experiments performed in Lund~[\ref{Lund}] for the 
$^{16}\mathrm{O}(\gamma,\mathrm{pn})^{14}\mathrm{N}$ reaction could separate 
final states, but, owing to low statistics, no momentum distribution for even 
the strongest states could be obtained. A new high resolution experiment 
for the $^{16}\mathrm{O}(\gamma,\mathrm{pn})^{14}\mathrm{N}$ and 
$^{16}\mathrm{O}(\gamma,\mathrm{pp})^{14}\mathrm{C}$ reactions in the 
photon-energy range between 90 and 270 MeV, aiming at separating final states 
and at determining also the momentum distributions for each state, has been 
recently approved in Mainz~[\ref{Mainz1}]. 

Various theoretical models have been developed over the last years to deal with 
two-nucleon emission processes. The model of ref.~[\ref{Oset}] aims at the 
best microscopic description of the photoabsorption process and includes many 
different reaction mechanisms. This approach is extremely useful to understand 
and disentangle the main mechanisms contributing in different kinematical 
regions and is able to reproduce their general features. However, it cannot 
describe the details of the nuclear structure aspects of the reaction, which 
are dealt with in a nuclear-matter approach with a local-density approximation. 
A detailed description of these aspects is essential if one wants to 
investigate the conditions of pairs of nucleons in finite nuclei and their 
state dependence through exclusive knockout reactions. A more proper 
description of nuclear structure properties is given by the two models 
developed in refs.~[\ref{GP},\ref{BGPR}] and~[{\ref{Ryck},\ref{Ryck1}], where, 
on the other hand, only the direct two-nucleon knockout mechanism is 
considered. Both of them include SRC, one-body and two-body currents and 
final-state interactions in a completely unfactorized calculation. These 
different aspects have been investigated at various steps and with somewhat 
different theoretical ingredients in the two models, which have both been 
applied to the analysis of existing data for electron- and photon-induced 
reactions. 

In refs.~[{\ref{Ryck},\ref{Ryck1}] reactions to discrete final states are not 
explicitly investigated but in very particular situations. In 
refs.~[\ref{GP},\ref{BGPR}] the various approximations used in the calculations 
make it difficult to evaluate the cross sections for transitions to individual 
final states. Thus a sum over all contributions given by different pairs of 
nucleons in the same shell is performed, which experimentally corresponds 
to detect all the pairs of nucleons coming out of the considered shell. This 
assumption is largely compatible with the energy resolution of the first 
experiments, but is not adequate for the analysis of the most recent exclusive 
data already available for the (e,e$'$pp) reaction~[\ref{Gerco}-\ref{Rosner}] 
and of the new data that are becoming available for both electron- and 
photon-induced reactions~[\ref{MAMI},\ref{Mainz1}]. 
                                  
Exclusive (e,e$'$pp) reactions have been investigated in 
refs.~[\ref{eepp},\ref{sf}]. In ref.~[\ref{eepp}] the theoretical model of 
refs.~[\ref{GP}] has been improved with a full treatment of antisymmetrization 
and of spin and isospin couplings. In ref.~[\ref{sf}] the pair removal 
amplitudes obtained from the calculation of the two-proton spectral function 
of $^{16} {\mathrm{O}}$~[\ref{Geurts}] have been included in the reaction 
calculation. The treatment of the $\Delta$ isobar current is more specifically 
discussed in ref.~[\ref{WAGP}], where some first numerical results for the 
exclusive ($\gamma$,pp) reaction are given but the spectral function is not 
included in the model. 

In this paper results for the exclusive 
$^{16}\mathrm{O}(\gamma,\mathrm{pn})^{14}\mathrm{N}$ and 
$^{16}\mathrm{O}(\gamma,\mathrm{pp})^{14}\mathrm{C}$ reactions to individual 
low-lying discrete final states of the residual nucleus are presented and 
discussed. These two reactions appear of particular interest. Exclusive data 
for both of them are expected from an approved experiment in 
Mainz~[\ref{Mainz1}]. 

The approach already applied to the 
$^{16}\mathrm{O(e,e'pp)}^{14}\mathrm{C}$ reaction~[{\ref{sf}] is here applied 
to the $^{16}\mathrm{O}(\gamma,\mathrm{pp})^{14}\mathrm{C}$ reaction. In this 
approach the pair removal amplitudes are obtained from the calculation of the 
spectral function, where both long-range and short-range correlations are 
treated consistently. Different relative and center-of-mass (CM) states of the 
pair contribute to a specific transition. They can be separated in the 
calculation and their effects can individually be investigated. A calculation 
of the spectral function for a pn pair is not available yet. Thus for the 
$^{16}\mathrm{O}(\gamma,\mathrm{pn})^{14}\mathrm{N}$ reaction the simpler 
prescription of ref.~[\ref{eepp}] is adopted here, where the two-nucleon 
overlap is given by the product of a coupled and fully antisymmetrized pair 
function of the shell model (SM) and a Jastrow type correlation function. In 
the two-body current for a pn pair contributions given by both seagull and 
pion-in-flight diagrams have been added to the $\Delta$ isobar current. This 
represents a further improvement with respect to the previous calculations for 
($\gamma$,pn) reactions of ref.~[\ref{BGPR}], where pion-in-flight diagrams 
were neglected. With respect to ref.~[\ref{Ryck1}], which is mainly devoted to 
investigate the quasi-deuteron kinematics and the behaviour of nucleon pairs 
knocked-out from different shells, this paper aims at investigating the 
transitions to discrete states, in order to explore their selectivity to the 
different components of the nuclear current. Moreover, the comparison of the 
results obtained with different models can give a deeper insight into the 
mechanisms of two-nucleon emission and the nuclear structure properties.

In this paper numerical results of cross sections and asymmetries are 
presented in the photon-energy range  between 100 and 400 MeV. It was already 
observed in ref.~[\ref{BGPR}] how the asymmetry of the cross section, that can 
be measured with linearly polarized photons, is helpful to investigate the 
reaction mechanism and the role of the different terms of the nuclear current. 
First asymmetry measurements for the $^{16}\mathrm{O}(\gamma,\mathrm{pp})$ and 
$^{16}\mathrm{O}(\gamma,\mathrm{pn})$ reactions have been carried out at 
LEGS~[\ref{LEGS}] in the photon energy range between 245 and 315 MeV. At these 
energies, around the peak of $\Delta$ resonance, the contribution of the isobar 
current is dominant and the calculations are extremely sensitive to its 
treatment. The data indicate the need of a careful description of this 
important theoretical ingredient. In this first experiment, however, the energy 
resolution was not sufficient to separate final nuclear states. 

The more refined treatment of spin degrees of freedom in the present approach 
is expected to give important effects on a polarization observable such as the 
asymmetry. It seems moreover interesting to investigate the dependence of this 
observable on the different states of the nucleon pair. 

The theoretical approach is outlined in sect.~2. Numerical results for the 
$^{16}\mathrm{O}(\gamma,\mathrm{pn})^{14}\mathrm{N}$ and 
$^{16}\mathrm{O}(\gamma,\mathrm{pp})^{14}\mathrm{C}$ reactions are 
presented and discussed in sect.~3 and sect.~4, respectively. Some conclusions 
are drawn in sect.~5.


\vspace{5mm}

\section*{2. Theoretical approach}
\baselineskip=0.75cm

\vspace{-5mm}

\hspace*{6mm}

\tightenlines
The coincidence cross section for the reaction induced by a photon, with energy 
$E_{\gamma}$, where two nucleons, with momenta $\p'_{1}$, and $\p'_{2}$ and 
energies $E'_{1}$ and $E'_{2}$, are ejected from a nucleus, is given, after 
integrating  over $E'_{2}$, by~[\ref{BGPR},\ref{Oxford}] 
\begin{equation}
\frac{{\mathrm d}^{5}\sigma}{{\mathrm d}E'_{1}{\mathrm d}\Omega'_{1} 
{\mathrm d}\Omega'_{2}} = \frac{\pi e^{2}}{2E_{\gamma}}\Omega_{\mathrm f} 
f_{\mathrm{rec}}^{-1} W_{\mathrm T} ,
\label{eq:cs}
\end{equation}
where $\Omega_{\mathrm f} = p'_{1} E'_{1} p'_{2} E'_{2}$ is the phase-space 
factor and integration over $E'_{2}$ produces the recoil factor 
\begin{equation}
f_{\mathrm{rec}}^{-1} = 1 - \frac{E'_{2}}{E_{\mathrm B}} \, \frac{\p'_{2}\cdot 
\p_{\mathrm B}}{|\p'_{2}|^2},
\end{equation}
where $E_{\mathrm B}$ and $\p_{\mathrm B}$ are the energy and momentum of the 
residual nucleus. The transverse structure function $W_{\mathrm T}$, which only 
depends on $E_{\gamma}$, $p'_{1}$, $p'_{2}$, the angles $\gamma_{1}$, between 
the momentum of the incident photon $\q$ and $\p'_{1}$, $\gamma_{2}$, between 
$\q$ and $\p'_{2}$, and $\gamma_{12}$, between $\p'_{1}$ and $\p'_{2}$, is 
expressed in terms of the components of the hadron tensor 
$W^{\mu\nu}$~[\ref{Oxford}], i.e.
\begin{equation}
W_{\mathrm T} = W^{xx} + W^{yy}
\end{equation}
and is thus given by bilinear combinations of the Fourier transforms 
of the transition matrix elements of the nuclear current density operator 
taken between initial and final nuclear states. 

If the residual nucleus is left in a discrete eigenstate of its Hamiltonian,
i.e. for an exclusive process, and under the assumption of a direct knockout
mechanism, the transition matrix elements can be written 
as~[\ref{GP},\ref{eepp}] 
\begin{eqnarray}
\J(\q) & = &  \int 
\psi_{\mathrm{f}}^{*}(\r_{1}\ss_{1},\r_{2}\ss_{2})
\J(\r,\r_{1}\ss_{1},\r_{2}\ss_{2})\psi_{\mathrm{i}}
(\r_{1}\ss_{1},\r_{2}\ss_{2}) \nonumber \\
& & \times \,{\mathrm{e}}^{\,{\mathrm{i}}{\footnotesize \q} \cdot
{\footnotesize \r}} {\mathrm d}\r{\mathrm d}\r_{1} {\mathrm d}\r_{2}
{\mathrm d}\ss_{1} {\mathrm d}\ss_{2} . \label{eq:jq}
\end{eqnarray}

Eq.~(\ref{eq:jq}) contains three main ingredients: the two-nucleon overlap
integral $\psi_{\mathrm{i}}$, the nuclear current $\J$ and the final-state
wave function $\psi_{\mathrm{f}}$. 

The derivation of Eq.~(\ref{eq:jq}) involves bound and scattering states,
$\psi_{\mathrm{i}}$ and $\psi_{\mathrm{f}}$, which are consistently
derived from an energy-dependent non-hermitian Feshbach-type Hamiltonian
for the considered final state of the residual nucleus. They are 
eigenfunctions of this Hamiltonian at negative and positive energy eigenvalues,
respectively~[\ref{GP},\ref{Oxford}]. In practice, it is not possible to 
achieve this consistency and the treatment of initial and final state 
correlations proceeds separately with different approximations.

In the final-state wave function $\psi_{\mathrm {f}}$ each of the outgoing
nucleons interacts with the residual nucleus while the mutual interaction
between the two outgoing nucleons is neglected. The scattering state is thus
written as the product of two uncoupled single-particle distorted wave
functions, eigenfunctions of a complex phenomenological optical potential
which contains a central, a Coulomb and a spin-orbit term. 

The two-nucleon overlap integral $\psi_{\mathrm{i}}$ contains the information 
on nuclear structure and allows one to write the cross section in terms of the 
two-hole spectral function~[\ref{Oxford}]. Since only a calculation of the 
two-proton spectral function of $^{16}{\mathrm {O}}$ is available at 
present~[\ref{Geurts},\ref{sf}], two different prescriptions have been here 
adopted for pp and pn knockout. 

For the $^{16}\mathrm{O}(\gamma,\mathrm{pp})^{14}\mathrm{C}$ reaction the 
two-nucleon overlap integrals are taken from the calculation of the 
spectral function~[\ref{Geurts},\ref{sf}]. They have been obtained from a 
two-step procedure, where long-range and short-range correlations 
are treated in a separate but consistent way. The calculation of long-range 
correlations is performed in a SM space large enough to incorporate 
the corresponding collective features which influence the pair removal 
amplitudes. The single-particle propagators used for this dressed Random 
Phase Approximation description of the two-particle propagator also include 
the effect of both long-range and short-range correlations. In the second step 
that part of the pair removal amplitudes which describes the relative motion 
of the pair is supplemented by defect functions obtained from the same 
G-matrix which is also used as the effective interaction in the RPA 
calculation.

For a discrete final state of the $^{14}$C nucleus, with angular momentum
quantum numbers $JM$, the two-nucleon overlap integral is expressed in terms 
of relative and CM wave functions as
\begin{equation}
\psi_{\mathrm{i}}(\r_{1}{\mbox{\boldmath $\sigma$}}_{1},
{\mbox{\boldmath $r$}}_{2}{\mbox{\boldmath $\sigma$}}_{2}) =  \sum_{nlSjNL} \,
c^{\,\mathrm{i}}_{nlSjNL} \, \phi_{nlSj}(r_{12}) R_{NL}(R) \,
\left[\Im ^{j}_{lS}
(\Omega_r,{\mbox{\boldmath $\sigma$}}_1,{\mbox{\boldmath $\sigma$}}_2) \,
Y_{L}(\Omega_R)\right]^{JM},
\label{eq:ppover}
\end{equation}
where   
\begin{equation}
\r_{12}= \r_{1}- \r_{2}, \, \, \, \, \,\R= \frac{(\r_{1} +\r_{2})}{2}
\end{equation}
are the relative and CM variables. The brackets in Eq.~({\ref{eq:ppover}) 
indicate angular momentum coupling of the angular and spin wave function $\Im$ 
of relative motion with the spherical harmonic of the CM coordinate to the 
total angular momentum quantum numbers $JM$. The CM radial wave function 
$R_{NL}$ is that of a harmonic oscillator (h.o.), with oscillator parameter 
$b=1.77$ fm. SRC are included in the radial wave function $\phi$ of relative 
motion through a defect function defined by the difference between $\phi$ 
and the uncorrelated relative h.o. wave function of the pair $R_{nl}$,
i.e.~[\ref{Geurts}]
\begin{equation}
\phi_{nlSj}(r_{12}) = R_{nl}(r_{12}) + D_{lSj}(r_{12}).
\label{eq:def}
\end{equation}
These defect wave functions were obtained by solving the Bethe-Goldstone 
equation in momentum space for $^{16}$O~[\ref{MS93a}]. They depend on 
different quantum numbers $l$, $S$, $l$. In the following the partial wave 
notation $^{2S+1}l_j$, for $l=S, P, D,$ is used for the relative states. 

The coefficients $c^{\,\mathrm{i}}$ in Eq.~(\ref{eq:ppover}) contain 
contributions from a SM space which includes the $0s$ up to the $1p0f$ 
shells. More details are given in ref.~[\ref{Geurts}] and in ref.~[\ref{sf}], 
where the same approach is applied to the reaction 
$^{16}\mathrm{O}(\mathrm e,\mathrm e '\mathrm{pp})^{14}\mathrm{C}$. 

For the reaction $^{16}\mathrm{O}(\gamma,\mathrm{pn})^{14}\mathrm{N}$, where a 
calculation of the spectral function is not available, the simpler 
prescription already applied in ref.~[{\ref{eepp}] to (e,e$'$pp) has been 
adopted here. The two-nucleon overlap is thus given by
\begin{equation}
\psi_{\mathrm{i}}(\r_{1}\ss_{1},\r_{2}\ss_{2}) \simeq
\Phi_{JM}(\r_{1}\ss_{1},\r_{2}\ss_{2})f(r_{12})
X_{TT_{3}}(\tau_{1},\tau_{2}),
\label{eq:over}
\end{equation}
where $\Phi_{JM} X_{TT_3}$ is the coupled and fully antisymmetrized
pair function of the shell model, $X_{TT_3}$is its isospin part and $f$ is a 
correlation function of Jastrow type which incorporates SRC. Only the central 
part of the correlation function is retained in the calculations. Thus 
long-range correlations are not included in this simpler approach.
The overlap function of Eq.~(\ref{eq:over}) describes a pair of nucleons
ejected from the same or different shells. The model includes the 
spectral strength $\lambda$. In the calculations we have assumed $\lambda = 1$. 
                  
The nuclear current operator in Eq.~(\ref{eq:jq}) is the sum of a one-body 
and a two-body part. In the one-body part convective and spin currents are 
included. The two-body current is derived from the effective Lagrangian of 
ref.~[\ref{Peccei}], performing a non relativistic reduction of the 
lowest-order Feynman diagrams with one-pion exchange. We have thus currents 
corresponding to the seagull and pion-in-flight diagrams and to the diagrams 
with intermediate isobar configurations, i.e.

\begin{eqnarray}
\J^{(2)}(\r,\r_{1}\ss_{1},\r_{2}\ss_{2}) & = &
\J^{\mathrm{sea}}(\r,\r_{1}\ss_{1},\r_{2}\ss_{2}) +
\J^{\pi}(\r,\r_{1}\ss_{1},\r_{2}\ss_{2}) \nonumber \\
& + & \J^{\Delta}(\r,\r_{1}\ss_{1},\r_{2}\ss_{2}) . \label{eq:nc}
\end{eqnarray}

In the coordinate space the seagull and pion-in-flight currents 
are~[\ref{Oxford}]
\begin{eqnarray}
\J^{\mathrm{sea}}(\r,\r_{1}\ss_{1},\r_{2}\ss_{2}) & = & - \frac{f^2}{4\pi}
\left(\tt_{1} \times \tt_{2}\right)_3 \, [\ss^{(1)}
\delta(\r_1-\r)(\ss^{(2)}\cdot \hat{\r}_{12})] \nonumber \\
& \times & \left(1 + \frac{1}{\mu r_{12}}\right) 
\frac{\mathrm{e}^{\,-\mu r_{12}}} {\mu r_{12}} + \, \, \, \, \, \, 
(1\leftrightarrow 2) ,
\label{eq:seag}
\end{eqnarray}
\begin{eqnarray}
\J^{\pi}(\r,\r_{1}\ss_{1},\r_{2}\ss_{2}) & = & - \frac{f^2}{16\pi^2}
\left(\tt_{1} \times \tt_{2}\right)_3 \, \dd_{1} 
\left(\ss_{1}\cdot\dd_{1}\right) \left(\ss_{2}\cdot\dd_{2}\right) \nonumber \\
& \times & \frac{\mathrm{e}^{\,-\mu |\r_1-\r|}}{\mu |\r_1-\r|} 
\frac{\mathrm{e}^{\,-\mu 
|\r_2-\r|}}{\mu |\r_2-\r|} + \, \, \, \, \, \,  (1\leftrightarrow 2) ,
\label{eq:pion}
\end{eqnarray}
where $f^2/(4\pi)=0.079$ and $\mu$ is the pion mass.

The operator form of the $\Delta$ current has been derived in 
ref.~[\ref{WAGP}]. It is given by the sum of the contributions of two types 
of processes, corresponding to the $\Delta$-excitation and 
$\Delta$-deexcitation currents. The first process (I) describes 
$\Delta$-excitation by photon absorption and subsequent deexcitation by pion 
exchange, while the second (II) describes the time interchange of the two 
steps, i.e., first excitation of a virtual $\Delta$ by pion exchange in a NN 
collision and subsequent deexcitation by photon absorption. The propagator of 
the resonance, $G_\Delta$, depends on the invariant energy $\sqrt s$ of the
$\Delta$, which is different for processes I and II. For the deexcitation 
current the static approximation can be applied, i.e. 
\begin{equation}
G_\Delta^{\,\mathrm{II}} = (M_\Delta-M)^{-1},    \label{eq:ei}
\end{equation}
where $M_{\Delta} = 1232$ MeV. For the excitation current we use~[\ref{WWA}]
\begin{equation}
G_\Delta^{\,\mathrm{I}} = \left({M_\Delta}-\sqrt{s_{\mathrm I}}-
\frac{{\mathrm{i}}}{2}\Gamma_\Delta (\sqrt{s_{\mathrm I}}) \right)^{-1},
\label{eq:prop}
\end{equation}
with 
\begin{equation}
\sqrt{s_{\mathrm I}}=\sqrt{s_{NN}}-M,
\label{eq:s1}
\end{equation}
where $\sqrt{s_{NN}}$ is the experimentally measured invariant energy
of the two outgoing nucleons and  the energy-dependent decay width of the 
$\Delta$, $\Gamma_\Delta$, has been taken in the calculations according to the 
parameterization of ref.~[\ref{BM}].
       
The sum of the two processes gives
\begin{eqnarray}
\J^{\Delta}(\r,\r_{1}\ss_{1},\r_{2}\ss_{2}) & = & \gamma \,
\delta(\r-\r_1)\,\{{\mathrm i} \,(G_\Delta^{\,\mathrm I} + 
G_\Delta^{\,\mathrm{II}}) \, [4\tau_{2,3} \A(\r_{12},\ss_{1},\ss_{2}) - 
\left(\tt_{1} \times \tt_{2}\right)_3 \nonumber \\
& \times & \B(\r_{12},\ss_{1},\ss_{2})] + 2 (G_\Delta^{\,\mathrm I}- 
G_\Delta^{\,\mathrm{II}}) \,  [\left(\tt_{1} \times 
\tt_{2}\right)_3 \A(\r_{12},\ss_{1},\ss_{2}) \nonumber \\
& + & \tau_{2,3}\B(\r_{12},\ss_{1},\ss_{2}) ]\}  \, \, + 
\, \, \, \, \, \, (1\leftrightarrow 2) ,
\label{eq:delta}
\end{eqnarray}
where                  
\begin{equation}
\A(\r_{12},\ss_{1},\ss_{2})  =  (\q \times \hat{\r}_{12}) \, (\ss_{2} \cdot
\hat{\r}_{12}) \, Y^{(1)}(r_{12}) \, - \, (\q\times\ss_{2}) \, 
Y^{(2)}(r_{12}), 
\end{equation}
\begin{equation}
\B(\r_{12},\ss_{1},\ss_{2})  =  \q\times (\ss_{1}\times \hat{\r}_{12}) \, 
(\ss_{2} \cdot \hat{\r}_{12}) \, Y^{(1)}(r_{12}) \, - \, \q\times\left(\ss_{1} 
\times \ss_{2}\right) \, Y^{(2)}(r_{12}), 
\end{equation}
\begin{eqnarray}
Y^{(1)}(r_{12}) & = &\left( 1 + \frac{3}{\mu r_{12}} + \frac{3}{\mu^2 
{r_{12}}^2}\right) \frac{\mathrm{e}^{\,-\mu r_{12}}}
{\mu r_{12}}, \\
Y^{(2)}(r_{12}) & = & \left( \frac{1}{\mu r_{12}} + \frac{1}{\mu^2 
{r_{12}}^2}\right) \frac{\mathrm{e}^{\,-\mu r_{12}}}
{\mu r_{12}}, 
\end{eqnarray}
and the factor $\gamma$ collects various coupling constants 
\begin{equation}
\gamma=\frac{f_{\gamma N\Delta}f_{\pi NN}f_{\pi  N\Delta}}{36\pi\mu}. 
\end{equation}

The charge-exchange term $\left(\tt_{1}\times\tt_{2}\right)_3$ in the two-body 
current vanishes for a pp pair. Thus only a part of the $\Delta$ current in 
Eq.~({\ref{eq:delta}) contributes to  ($\gamma$,pp) reactions, whereas all the 
terms in eqs.~({\ref{eq:seag},\ref{eq:pion},\ref{eq:delta}) contribute to 
($\gamma$,pn). On this basis it is generally expected that the two-body 
current is much more important for the knockout of a pn pair than of a pp pair. 

From the analysis of the isospin matrix structure of processes I and II, it 
can be understood~[\ref{WAGP}] that for photon absorption on a pn pair 
only the excitation current contributes in a state with $T=0$, whereas only the 
deexcitation current contributes in a state with $T=1$. Since the propagator 
$G_\Delta^{\,\mathrm I}$ gives rise to a pronounced resonant behaviour of the 
matrix elements of the excitation current in the photon-energy region between 
200 and 400 MeV, we expect that in this region the contribution of the 
$\Delta$ current for the knockout of a pn pair with $T=0$ is much larger than 
for a pn pair with $T=1$. 

Both processes I and II contribute to photon absorption on a pp pair ($T=1$). 
For the  knockout of a pp pair in a relative $^1S_0$ state the 
generally dominant magnetic dipole $NN \leftrightarrow N\Delta$ transition is 
suppressed because of total angular momentum and parity 
conservation~[{\ref{WNA}]. In our calculations contributions of all multipoles 
are included through the term ${\mathrm e}^{\,{\mathrm i}{\footnotesize \q} 
\cdot {\footnotesize \r}}$, which represents the incident photon in the 
transition amplitudes of Eq.~(\ref{eq:jq}). Only if we set 
${\mathrm e}^{\,{\mathrm i}{\footnotesize \q} \cdot {\footnotesize \r}} = 1$ 
the calculation for the $\Delta$ current is restricted to a magnetic-dipole 
transition. In any case the effect of higher multipoles is generally smaller 
and for a $^1S_0$ pp pair the $\Delta$ current, although nonvanishing, turns 
out to be much less important than for other relative states. This was already 
observed in ref.~[{\ref{sf}] for the (e,e$'$pp) reaction. In that case the 
contribution of the excitation current, which is generally dominant, is 
strongly reduced on a $^1S_0$ pp pair, where it becomes of about the same size 
or even smaller than that of the deexcitation current, which is generally 
small. As a consequence, in (e,e$'$pp) reactions situations where the removal 
of a $^1S_0$ pair is dominant are particularly well suited to emphasize and 
thus investigate effects of SRC~[\ref{Gerco},\ref{sf}]. This conclusion cannot 
be directly applied to ($\gamma$,pp) reactions, where only the transverse 
component of the nuclear response contributes and the longitudinal component, 
where the effects od SRC show up more strongly, is absent. 

In photon-induced reactions with linearly polarized photons one can measure 
the photon asymmetry $\Sigma$, given by the difference between the cross 
sections with linear photon polarization parallel and perpendicular to the 
reaction plane. It can also be expressed in terms of  the ratio between the 
two transverse structure functions $W_{TT}$ and 
$W_{T}$~[\ref{BGPR},\ref{Oxford}], as
\begin{equation}
\Sigma = \frac{W^{xx}-W^{yy}} {W^{xx}+W^{yy}} = \frac{W_{TT}}{W_{T}}.
\label{eq:asym}
\end{equation}

The asymmetry is particularly sensitive to spin variables and thus to the 
different terms of the nuclear current and to their 
interference, whose effects on $W_{T}$ and $W_{TT}$ are emphasized in the 
ratio, where, on the contrary, other effects are smoothed 
out~[\ref{BGPR},\ref{Ryck1}].                  

In the following numerical results are presented of cross sections and 
asymmetries for the $^{16}\mathrm{O}(\gamma,\mathrm{pn})^{14}\mathrm{N}$ and 
$^{16}\mathrm{O}(\gamma,\mathrm{pp})^{14}\mathrm{C}$ reactions for transitions
to discrete final states of the residual nucleus. High-resolution experiments 
are needed to separate individual final states in the excitation-energy 
spectrum. In our calculations each state is characterized by a particular 
value of the missing energy, given by 
\begin{equation}
E_{2\mathrm{m}} = E_\gamma - T'_{1} - T'_{2} -T_{\mathrm{B}} =
E_{\mathrm{s}} + E_{\mathrm{x}} , \label{eq:em}
\end{equation}
where $T'_{1}$, $T'_{2}$ and $T_{\mathrm{B}}$ are the kinetic energies of the
two outgoing nucleons and of the residual nucleus, respectively,
$E_{\mathrm{s}}$ is the separation energy at threshold for two-nucleon
emission and $E_{\mathrm{x}}$ is the excitation energy of the residual nucleus.


\vspace{5mm}

\section*{3. The reaction 
\boldmath{$^{16}\mathrm{O}(\gamma,\mathrm{\lowercase{pn}})^{14}\mathrm{N}$}}
\baselineskip=0.75cm

\vspace{-5mm}

\hspace*{6mm}

\tightenlines
The theoretical approach for two-nucleon knockout outlined in sect.~2 has been 
applied to the reaction $^{16}\mathrm{O}(\gamma,\mathrm{pn})^{14}\mathrm{N}$}, 
for transitions to discrete final states of the residual nucleus. Only the 
states with an excitation energy lower than 12 MeV have been considered in 
the calculations, according to the experimental analysis of 
ref.~[\ref{Mainz}], where the direct two-nucleon knockout mechanism is well 
established for missing energies up to 40 MeV in $^{12}\mathrm{C}$. Therefore 
we consider here transitions to the 1$^+_1$ ($T=0$) ground state of 
$^{14}\mathrm{N}$, to the 0$^+$ ($T=1$) state at 2.31 MeV, to the 1$^+_2$ 
($T=0$) state at 3.95 MeV, to the 2$^+$ ($T=0$) state at 7.03 MeV and to the 
3$^+$ ($T=0$) state at 11.05 MeV. We  do not consider the negative parity 
states at low energy and other positive parity states, as they are not assumed 
as two-hole states in the independent particle model and therefore are not 
likely to be excited in a direct knockout~[\ref{C14}].

The two-nucleon overlap is taken in the calculations according to the 
prescription of Eq.~(\ref{eq:over}). In the SM wave function, all the 
considered states are described as two-hole states in the $p$ shell: 1$^+_1$ 
and 0$^+$ are ($p_{1/2}$)$^{-2}$ holes, 1$^+_2$ and 2$^+$ are 
($p_{1/2}p_{3/2}$)$^{-1}$ holes and 3$^+$ is a ($p_{3/2}$)$^{-2}$ hole. In the 
frame of a h.o. basis, their different components in the relative and CM 
motion are given in Table I with the corresponding amplitudes. This 
decomposition can be helpful to understand the general features of cross 
sections and asymmetries. In the calculations, however, the more realistic 
Woods-Saxon single-particle wave functions of ref.~[\ref{ES}] have been used. 
The correlation function is taken from ref.~[\ref{GD}]. 

The distorted wave functions in the final state are eigenfunctions of the 
optical potential of ref.~[\ref{Nad}], which includes also the spin-orbit 
component. 

As explained in sect.~2, the nuclear current is the sum of a one-body part, 
whose contribution is entirely due to the correlation function, and a two-body 
part, including contributions of the lowest-order diagrams with one-pion 
exchange, i.e. seagull, pion-in-flight and diagrams with intermediate isobar 
configurations. 

Calculations have been performed in the coplanar symmetrical kinematics,
where the two nucleons are emitted in the scattering plane at equal energies
and equal but opposite angles with respect to the beam direction, i.e. with 
$\gamma_1=\gamma_2$ and $\gamma_{12}=\gamma_1+\gamma_2$. In this kinematics, 
if $E_\gamma$ is fixed and $\gamma_1$ is varied, it is possible to explore, 
for different scattering angles, all possible values of the recoil 
($p_{\mathrm{B}}$) or missing momentum ($p_{2\mathrm{m}}$) distribution, where 
\begin{equation}
\p_{2\mathrm{m}} = \p_{\mathrm{B}} =  \q -\p'_{1} - \p'_{2}.  \label{eq:pm}
\end{equation}
If relative and CM motion are factorized and final-state interactions are 
neglected~[\ref{Gottfried}], $\p_{\mathrm{B}}$ is opposite to the total 
momentum $\P$ of the nucleon pair in the target. 

The choice of this kinematics is made at variance with that of 
ref.~[\ref{Ryck1}], where particular attention is drawn to the dynamics of 
dinucleons in the medium, which is studied through a kinematics where 
$\p_{\mathrm{B}}$ equals zero. The symmetrical kinematics is best suited to 
give information on the motion of the pair in different CM and relative 
angular momentum states and therefore to evaluate the effects of the one-body 
and two-body components of the nuclear current on these states, which is a 
goal of this paper.

The calculated cross sections allow us to investigate a missing-momentum range 
up to about 250-300 MeV/$c$ on both sides of the distributions. A wide range 
of photon energies, between 100 and 400 MeV, has been explored.

The angular distributions corresponding to the five considered final states 
are given at E$_\gamma$=100 MeV in Figs.~1 and 2 and at E$_\gamma$=300 MeV in 
Figs.~3 and 4. In the figures separate contributions of the different terms of 
the nuclear current are compared with the cross sections produced by the sum 
of all of them. Moreover also the sum of the seagull and pion-in-flight 
currents and their sum with the one-body current are drawn, in order to better 
evaluate interference effects. The dependence on the photon energy in the 
considered range is given in Figs.~5 and 6, for the different contributions, 
at a fixed value of $\gamma_1$ near to the maximum of the distribution of each 
state. 

The shape of the angular distributions in Figs.~1-4 is determined by the 
combined effect of the CM wave function of the pair and of the different terms 
of the nuclear current. The main effect is given by the components of the CM 
wave function contributing to each state: 0$^+$ and 1$^+_2$ have essentially 
an $s$-wave shape, 1$^+_1$ is due to a combination of $p$, $d$ and $s$ waves, 
while 2$^+$ and 3$^+$ are a combination of $d$ and $s$ waves at $E_\gamma=100$ 
MeV and have a $d$-wave shape at $E_\gamma=300$ MeV, where the contribution of 
the $\Delta$ current is dominant. This analysis is consistent with the 
decomposition presented in Table I, for h.o. shell model wave 
functions, combined with the different values of the nuclear current matrix 
elements of the different relative states. 

The results in Figs.~1-4 indicate that the contribution of the one-body 
current and thus of SRC on the cross section is generally extremely small. It 
increases with the photon energy but is always overwhelmed by the two-body 
current. It is very sensitive to the choice of the correlation function: the 
strong correlation function of ref.~[\ref{Clark}] calculated with the 
hard-core OMY NN--interaction~[\ref{OMY}], for instance, can enhance the 
contribution of the one-body current by about an order of magnitude. Even in 
this case, however,  the calculated cross sections are dominated by two-body 
currents. The use in the present approach for pn knockout of a central and 
state-independent correlation function represents a very simple description of 
SRC. Moreover these correlation functions are not specifically calculated for 
a pn pair. Nevertheless, this simple treatment should be able to give a 
reasonable prediction of the role of SRC in the considered situations. Our 
numerical results indicate that ($\gamma$,pn) cross sections are dominated by 
two-body currents and a more sophisticated  treatment of correlations should 
not change this conclusion. This result is in agreement with that of 
refs.~[\ref{Ryck1},\ref{LEGS}].

The seagull current is dominant, for all the considered transitions, at 
$E_\gamma =100$ MeV and in general at low values of the photon energy, up to 
about 150 MeV. Its pure contribution slightly decreases when the photon energy 
increases. The contribution of the pion-in-flight current is always much 
smaller than that of the seagull current and is only weakly dependent on the 
photon energy. It was found in refs.~[\ref{Ryck},\ref{Ryck1}] that a 
destructive interference is given by the sum of these two terms. This result 
is confirmed by the present calculations. However here this effect is 
generally small, but for the 0$^+$ state, i.e. the only state with $T=1$, 
where a strong destructive interference is obtained. 

The isobar current shows, as expected, a strong dependence on the photon 
energy. Its contribution is negligible at low values of $E_\gamma$, as in 
Figs.~1 and 2, but gains importance as the energy increases and becomes 
dominant above 200 MeV. For the four states with $T=0$ only the 
$\Delta$-excitation current with its resonant propagator contributes. The 
resonance peak is clearly shown in the energy distributions of Figs.~5-6. On 
the contrary, only the $\Delta$-deexcitation current contributes for the 
0$^+$ ($T=1$) state. Indeed for this state the $\Delta$ current does not show a 
resonant behaviour and above the pion-production threshold its contribution, 
although still dominant in the photon-energy range between 250 and 400 MeV, 
is smaller than for the other states. 

In Figs.~7-10 the photon asymmetries are shown in the same situations and in 
the same conditions as in Figs.~1-4 for the cross section. Different shapes 
are obtained for different states. The asymmetry is sensitive to the nuclear 
current, whose different effects on the two structure functions $W_{T}$ and 
$W_{TT}$ in Eq.~(\ref{eq:asym}) are emphasized in the ratio. The differences 
between the results in Figs.~7 and 8, at $E_\gamma=100$ MeV,  and in Figs.~9 
and 10, at $E_\gamma=300$ MeV, indicate the role of the different terms of the 
nuclear current at the two values of the photon energy. 

The asymmetry given by the contribution of the various terms of the current 
does not show a strong dependence on the photon energy. Like the cross 
section, it is not significantly affected by the one-body current. On the 
contrary, as it was already observed in the different model of 
ref.~[\ref{Ryck1}], it is very sensitive to all the components of the two-body 
current and to their interference. The pure contributions of the seagull and 
pion-in-flight currents generally give a positive asymmetry. In particular, 
for the seagull current it is positive and large (close to 1) for the 0$^+$ 
state and positive and small (around 0) for the 2$^+$ state. A large effect is 
given by the interference between seagull and pion-in-flight terms, which 
generally produces a negative asymmetry. Only for the $0^+$ state the 
asymmetry given by the sum of the two terms remains positive. This strong 
interference effect is essential to produce the final result at low values of 
$E_\gamma$, e.g. in Figs.~7 and 8 at $E_\gamma=100$ Mev. The contribution of 
the pion-in-flight terms and of its interference with the seagull current is 
therefore much more important for the asymmetry than for the cross section. 
The asymmetry produced by the $\Delta$ current is generally negative for all 
the considered states. In the calculations, however, it presents anomalies and 
sharp peaks corresponding to minima of the cross section. At large values of 
the photon energy, e.g. in Figs.~9 and 10 at 300 MeV, the asymmetry, like the 
cross section, is dominated by the $\Delta$ current. However, for the $0^+$ 
state, where only the deexcitation part of the isobar current contributes, the 
effect of the other terms  of the two-body current is relatively a bit more 
important in determining the final result. 


\vspace{5mm}

\section*{4. The reaction \boldmath{$^{16}\mathrm{O}(\gamma,
\mathrm{\lowercase{pp}})^{14}\mathrm{C}$}}
\baselineskip=0.75cm

\vspace{-5mm}

\hspace*{6mm}

\tightenlines
For the reaction $^{16}\mathrm{O}(\gamma,\mathrm{pp})^{14}\mathrm{C}$ 
the theoretical approach of sect.~2 has been applied to the transitions to the 
three lowest discrete final states of the residual nucleus, i.e. the 0$^+$ 
ground state of $^{14}\mathrm{C}$, the 2$^+$ state at 7.01 MeV and the 
1$^+$ state at 11.31 MeV. These are the states that have already been 
experimentally separated in the $^{16}$O(e,e$'$pp)$^{14}$C 
reaction~[\ref{Gerco},\ref{Gercoth},\ref{Rosner}]. 

The two-nucleon overlap functions for each state are taken in the calculations 
from the spectral function of refs.~[\ref{Geurts},\ref{sf}]. Their 
explicit expressions, Eq.~(\ref{eq:ppover}), allow the separation of the 
different components of relative and CM motion. In Table II the main 
components and their amplitudes are given and compared with the corresponding 
amplitudes obtained, with h.o. wave functions, in the SM, where the states are 
described as two holes in the $p$ shell: 0$^+$ is a ($p_{1/2}$)$^{-2}$ hole,
while 2$^+$ and 1$^+$ are ($p_{1/2}p_{3/2}$)$^{-1}$ holes. While the SM 
amplitudes are normalized to one, the spectral density amplitudes are 
normalized to the spectroscopic factor, i.e. to about 0.6. The table shows 
that in the spectral function the amplitudes for the $^3P_j$ relative states 
are more reduced than for the $^1S_0$ states with respect to the corresponding 
SM amplitudes. Other components, not shown in the table, contribute to the 
spectral density function and are included in the calculation, but their 
contribution is less than 1\%~[\ref{Geurts},\ref{sf}].

The same conditions and the same ingredients used for the ($\gamma$,pn) 
reaction in sect.~3 have been adopted. For a pp pair, 
however, the two-body part of the nuclear current does not include the 
charge-exchange terms and contains only the non-exchange component of the
$\Delta$ isobar current.

In order to explore the missing-momentum distribution of each state, the same 
coplanar symmetrical kinematics as in sect.~3 has been considered, with the 
same range of photon energies and scattering angles.

The angular distributions corresponding to the considered final states 
of $^{14}$C are given in Figs.~11 and 12, at $E_\gamma=100$ MeV, and in 
Fig.~13, at $E_\gamma=300$ MeV. In Fig.~11 the separate contributions of the 
one-body and $\Delta$ currents are compared with the cross sections given by 
the sum of the two terms. In Figs.~12 and 13 the contributions of the 
different relative states are drawn. 

The shape of the angular distributions is determined by the CM orbital angular 
momentum of the knocked-out pair. The 0$^+$ state has essentially an $s$-wave 
shape; the $p$-wave component becomes more important at large values of the 
missing momentum, where the $s$-wave contribution becomes small. The 1$^+$ 
state has a $p$-wave shape and the 2$^+$ state is a combination of $p$ and $d$ 
waves. 

The $\Delta$ current is dominant in the cross section, for all the considered 
transitions and photon-energy values. Its contribution strongly increases 
with the energy, and above the pion-production threshold shows a pronounced 
resonant behaviour, with a maximum at the peak of the resonance. It is reduced 
in the $^1S_0$ state, where, as explained in sect.~2, the magnetic dipole 
$NN \leftrightarrow N\Delta$ transition is forbidden. 

The contribution of the one-body current and therefore the effect of SRC is 
generally very small. SRC are accounted for in the calculations by defect 
functions which depend on the relative states. The results of Figs.~11-13 
have been obtained with the defect functions calculated with the Bonn-A 
potential~[\ref{Geurts}]. A different choice sensibly changes the contribution 
of the one-boby current, but does only sligthly influence the final result, 
which is driven by the $\Delta$ current. An example is shown in Fig.~14, where 
the cross sections obtained with the defect functions from Bonn-A and Reid 
Soft Core potentials~[\ref{Geurts}] are compared, for the 0$^+$ and 2$^+$ 
states, at $E_\gamma=100$ MeV where the contribution of the $\Delta$ current 
is smaller. At higher energies the one-body contribution increases and the 
results with Reid defect functions become larger than with Bonn-A. The 
increase with the energy of the contribution of the $\Delta$ current is 
however much larger and the final cross sections are even more dominated by 
the two-body current. Our results indicate that also ($\gamma$,pp) cross 
sections are only slightly sensitive to SRC in the considered situations.
 
In Fig. 15 the photon asymmetries are shown for the 1$^+$ and 2$^+$ states at 
$E_\gamma$ = 100 and 300 MeV. The contributions of the different relative 
states are shown with the same convention as in Figs.~12 and 13. For the 0$^+$ 
state, not drawn in the figure, the asymmetry is always negative and almost 
everywhere equal to $-$1. This result was already observed in 
ref.~[\ref{Ryck1}].

The asymmetry, like the cross section, is dominated by the $\Delta$ current and 
does  not show a strong dependence on the photon energy. Different results 
are obtained for different states. For 2$^+$ the asymmetry is negative, while 
for 1$^+$ it is large and positive, in agreement with the results of 
ref.~[\ref{Ryck1}], due to the large positive value (about 1) produced by the 
$\Delta$ current in the $^3P_0$ state, which gives the dominant contribution 
to the final result. 

The asymmetry is not significantly affected by the one-body current. In  
Fig.~16 the effect of the different choices of the defect functions from 
Bonn-A and Reid potentials is shown for the 2$^+$ state at $E_\gamma=100$ MeV. 
It appears that the choice of the defect functions can have a significant but 
not substantial effect on the asymmetry, confirming that ($\gamma$,NN) 
reactions are not particularly well suited for studying SRC.


\vspace{5mm}

\section*{5. Summary and conclusions}
\baselineskip=0.75cm

\vspace{-5mm}

\hspace*{6mm}

\tightenlines
In this paper exclusive 
$^{16}\mathrm{O}(\gamma,\mathrm{pn})^{14}\mathrm{N}$ and 
$^{16}\mathrm{O}(\gamma,\mathrm{pp})^{14}\mathrm{C}$ knockout reactions have 
been investigated, for transitions to the low-lying discrete final states of 
the residual nucleus.

Exclusive ($\gamma$,NN) reactions may represent a stringent test of reaction 
mechanisms and a promising tool for investigating the dynamics of bound pairs 
of nucleons in the nuclear medium. The presence of several discrete final 
states with well-defined angular momentum makes the $^{16}$O nucleus an 
excellent target for this analysis. Measurements with sufficient energy 
resolution to separate individual final states and good statistical accuracy 
to determine for each state the dependence of the cross sections on dynamical 
variables are needed. These data are now becoming available from new 
high-resolution experiments~[\ref{Mainz1}]. 

Numerical predictions of cross sections and photon asymmetries have been 
presented in the photon-energy range between 100 and 400 MeV. These results
have been obtained in a direct knockout framework where both one-body and 
two-body currents are included. The contribution of the one-body current is 
entirely due to SRC; the two-body terms include contributions of the 
lowest-order Feynman diagrams with one-pion exchange, i.e. seagull, 
pion-in-flight and diagrams with intermediate isobar configurations, where a 
$\Delta$ is excited or a preformed $\Delta$ is deexcited after the absorption 
of the photon. Final-state interactions are taken into account by means of a 
phenomenological spin-dependent optical potential describing the interaction 
of each of the outgoing nucleons with the residual nucleus. The two-nucleon 
overlap functions have been obtained for the 
$^{16}\mathrm{O}(\gamma,\mathrm{pp})^{14}\mathrm{C}$ reaction from a recent 
calculation of the spectral function~[\ref{Geurts}], where both long-range and 
short-range correlations are consistently treated. This approach was already 
applied to the $^{16}\mathrm{O(e,e'pp)}^{14}\mathrm{C}$ reaction~[{\ref{sf}]. 
The two-nucleon overlap functions for the 
$^{16}\mathrm{O}(\gamma,\mathrm{pn})^{14}\mathrm{N}$ reaction are given by the 
product of a coupled and fully antisymmetrized SM pair function and a Jastrow 
type correlation function, which incorporates SRC, since a calculation of the 
spectral function for a pn pair is not available. Although presumably too 
simplistic, this prescription should be able to give reasonable predictions of 
the role of SRC in ($\gamma$,pn) reactions.

The angular distributions of the cross sections for transitions to different 
states have different shapes, essentially determined by the components of the 
CM orbital angular momentum of the initial pair. Also the asymmetry is 
sensitive to the behaviour of the different components of the pair wave 
functions. These results confirm that exclusive reactions are particularly 
well suited to check the relevance of the direct knockout mechanism and to 
explore the conditions of different pairs of nucleons in the nuclear medium. 

Cross sections and asymmetries, for both ($\gamma$,pn) and ($\gamma$,pp) 
reactions, are only slightly affected by SRC and are dominated, in all the 
considered situations, by two-body currents. Therefore photon-induced 
two-nucleon knockout reactions do not seem particularly well suited to 
investigate SRC, but might give interesting information on the various terms 
of the nuclear current and on their behaviour in different kinematic 
conditions and in different states. 

Better information on correlations for a pn pair might be obtained from 
the (e,e$'$pn) reactions, where also the longitudinal component of the nuclear 
response contributes. In this case, as it was already observed in 
(e,e$'$pp)~[\ref{sf}], a suitable choice of kinematics in exclusive 
experiments might allow the separation of effects due to SRC and two-body 
currents. For this analysis, however, a more refined treatment of correlations 
would be desirable, such as that applied here to pp pairs through  the 
spectral function. Moreover, also tensor correlations might be investigated 
for nucleon pairs in a $T=0$ state. 

Our results indicate that in the 
$^{16}\mathrm{O}(\gamma,\mathrm{pn})^{14}\mathrm{N}$ reaction the 
seagull current is dominant at lower values of the photon energy, up to about 
150 MeV, then the $\Delta$ current gains importance. The contribution of 
the pion-in-flight current is always much smaller than that of the seagull 
current and is only weakly dependent on the photon energy. In the cross 
sections it generally produces a destructive interference with the seagull 
currents. This effect is however small, but for the 0$^+$ ($T=1$) state. 
Interference effects are better emphasized in the photon asymmetry, where the 
interference between seagull and pion-in flight terms turns the asymmetry from 
positive, for both separate contributions, to negative, but for the $0^+$ 
state. This effect is essential at low values of the photon energy, while at 
higher energies both cross sections and asymmetries are dominated by the 
$\Delta$ current. The $\Delta$ current is dominant also in the 
$^{16}\mathrm{O}(\gamma,\mathrm{pp})^{14}\mathrm{C}$ reaction, for all the 
considered transitions and over all the considered range of photon energy. Its 
contribution strongly increases with the energy, but plays the main role, both 
on cross sections and asymmetries, even at 100 MeV. 

The results of our investigation indicate that a good description of the 
isobar current is indispensable in the theoretical treatment. On the other 
hand, ($\gamma$,NN) reactions might give deeper insight into the behaviour of 
the $\Delta$ in the nuclear medium. Moreover, also the influence of two-body 
currents due to heavier meson exchange could deserve further investigation.\\

We are grateful to K. Allaart and P. Grabmayr for useful comments and fruitful
discussions.

\newpage
    
\section*{References}
\tightenlines
\newcounter{ref}
\begin{list}
{[\arabic{ref}]}{\usecounter{ref}}\setlength{\rightmargin}{\leftmargin}

\item \label{Gottfried}
K. Gottfried, Nucl. Phys.  5 (1958) 557; Ann. of Phys.  21
(1963) 29; \newline W. Czyz and K. Gottfried Ann. of Phys. 21 (1963) 29; 
\newline 
Y.N. Srivastava,  Phys. Rev.  135 (1964) B612; \newline
D.U.L. Yu, Ann. of Phys.  38 (1966) 392.
\vspace{-3mm}

\item \label{Oxford}
S. Boffi, C. Giusti, F.D. Pacati and M. Radici, Electromagnetic response of
atomic nuclei, Oxford Studies in Nuclear Physics (Clarendon Press,
Oxford, 1996).
\vspace{-3mm}

\item \label{Kester}
L.J.H.M. Kester, W.H.A. Hesselink, N. Kalantar-Nayestanaki, J.H. Mitchell, A. 
Pellegrino, E. Jans, J. Konijn, J.J.M. Steijger, J.L. Visschers, A. Zondervan, 
J.R. Calarco, D. DeAngelis, F.W. Hersman, W. Kim, Th.S. Bauer, M.W. Kelder, Z. 
Papandreou, C. Giusti and F.D. Pacati, Phys. Rev. Lett.  74 (1995) 1712.
\vspace{-3mm}
  
\item \label{Zondervan}
A. Zondervan, L.J. de Bever, E. Jans, J. Konijn, M. Kruijer, J.J.M. Steijger, 
J.L. Visschers, P.J. Countryman, W.H.A. Hesselink, N. Kalantar-Nayestanaki, 
L.J.H.M. Kester, J.H. Mitchell, A. Pellegrino, J.R. Calarco, F.W. Hersman, M. 
Leuschner, T.P. Smith, Th.S. Bauer, M.W. Kelder, C. Giusti, F.D. Pacati, J. 
Ryckebusch and M. Vanderhaeghen, Nucl. Phys.  A 587 (1995) 697.
\vspace{-3mm}

\item \label{GP}
C. Giusti and F.D. Pacati, Nucl. Phys.  A 535 (1991) 573; Nucl. Phys.  A 571 
(1994) 694; Nucl. Phys.  A 585 (1995) 618.
\vspace{-3mm}

\item \label{BGPR}
C. Giusti, F.D. Pacati and M. Radici, Nucl. Phys.  A 546 (1992) 607;\newline
S. Boffi, C. Giusti, F.D. Pacati and M. Radici, Nucl. Phys.  A 564 
(1993) 473.
\vspace{-3mm}

\item \label{Gerco}
C.J.G. Onderwater, K. Allaart, E.C. Aschenauer, Th.S. Bauer, D.J. Boersma, E. 
Cisbani, S. Frullani, F. Garibaldi, W.J.W. Geurts, D.L. Groep, W.H.A. 
Hesselink, M. Jodice, E. Jans,  N. Kalantar-Nayestanaki, W.-J. Kasdorp, C. 
Kormanyos, L. Lapik\'as, J.J. van Leeuwe, R. de Leo, A. Misiejuk, A.R. 
Pellegrino, R. Perrino, R. Starink, M. Steenbakkers, G. van der Steenhoven, 
J.J.M. Steijger, M.A. van Uden, G.M. Urciuoli, L.B. Weinstein and H.W. 
Willering, Phys. Rev. Lett.  78 (1997) 4893.
\vspace{-3mm}

\item \label{Gercoth}
C.J.G. Onderwater, Ph.D. thesis,  Vrije Universiteit, Amsterdam (1998).
\vspace{-3mm}

\item \label{Rosner}
G. Rosner, Conference on Perspectives in Hadronic Physics, eds. S. Boffi, C. 
Ciofi degli Atti, and M.M. Giannini (World Scientific, Singapore), in press.
\vspace{-3mm}

\item \label{MAMI}
P. Bartsch, D. Baumann, R. B{\"o}hm, T. Caprano, S. Derber, M. Ding, A. Ebbes, 
I. Ewald, J. Friedrich, J.M. Friedrich, R. Geiges, P. Jennewein, M. Kahrau, M. 
Korn, K.W. Krygier, A. Liesenfeld, H. Merkel, K. Merle, P. Merle, U. 
M{\"u}ller, R. Neuhausen, Th. Pospischil, G. Rosner (spokesperson), H. 
Schmieden, A. Wagner, Th. Walcher, M. Weis, S. Wolf, MAMI proposal Nr: A1/1-97.
\vspace{-3mm}

\item \label{eepp}
C. Giusti and F.D. Pacati, Nucl. Phys.  A 615 (1997) 373.
\vspace{-3mm}

\item \label{sf}
C. Giusti, F.D. Pacati, K. Allaart, W.J.W. Geurts, W.H. Dickhoff and H. 
M{\"u}ther, Phys. Rev. C 57 (1998) 1691. 
\vspace{-3mm}

\item \label{Geurts}
W.J.W. Geurts, K. Allaart, W.H. Dickhoff and H. M{\"u}ther, Phys. Rev.  C 54 
(1996) 1144.
\vspace{-3mm}

\item \label{Ryck}
J. Ryckebusch, L. Machenil, M. Vanderhaengen, and M. Waroquier, Phys. Lett. 
B 291 (1992) 213; \newline 
L. Machenil, M. Vanderhaengen, J. Ryckebusch, and M. Waroquier, Phys. Lett. 
B 316 (1993) 17; \newline 
J. Ryckebusch, L. Machenil, M. Vanderhaengen, V. Van der Sluys and M. 
Waroquier, Phys. Rev.  C 49 (1994) 2704; \newline 
J. Ryckebusch, L. Machenil, M. Vanderhaengen, and M. Waroquier, Nucl. Phys. 
A 568 (1994) 828; \newline 
M. Vanderhaengen, L. Machenil, J. Ryckebusch, and M. Waroquier, Nucl. Phys. 
A 580 (1994) 551.
\vspace{-3mm}

\item \label{Ryck1}
J. Ryckebusch,  D. Debruyne and W. Van Nespen, Phys. Rev.  C 57 (1998) 1319.
\vspace{-3mm}

\item \label{Bonn}
J. Arends, J. Eyimk, H. Hartmann, A. Hegerath, B. Mecking, G. N{\"o}ldeke, and 
H. Rost,  Z. Phys. (1980) 103.
\vspace{-3mm}

\item \label{Tokio}
M. Kanazawa, S. Homma, M. Koike, Y. Murata, H. Okuno, F. Soga, N. Yoshikawa and 
A. Sasaki, Phys. Rev. C 35 (1987) 1828.
\vspace{-3mm}

\item \label{MacG}
J.C. McGeorge, I.J.D. MacGregor, S.N. Dancer, J.R.M. Annand, I. Anthony, G.I. 
Crawford, S.J. Hall, P.D. Harty, J.D., Kellie, G.J. Miller, R.O. Owens, P.A. 
Wallace, D. Branford, A.C. Shotter, B. Schoch, R. Beck, H. Schmieden and J.M. 
Vogt, Phys.  Rev. C 51 (1995) 1967.
\vspace{-3mm}

\item \label{Mainz}
P. Grabmayr, J. Ahrens, J.R.M. Annand, I. Anthony, D. Branford, G.E. Cross, T. 
Davinson, S.J. Hall, P.D. Harty, T. Hehl, J.D. Kellie, Th. Lamparter, I.J.D. 
MacGregor, J.A. MacKenzie, J.C. McGeorge, G.J. Miller, R.O. Owens, M. Sauer, 
R. Schneider, K. Spaeth and G.J. Wagner, Phys. Lett. B 370 (1996) 17;\newline
P.D. Harty, I.J.D. MacGregor, P. Grabmayr, J. Ahrens, J.R.M. Annand, I. 
Anthony, R. Beck, D. Branford, G.E. Cross, T. Davinson, S.J. Hall, T. Hehl, 
J.D. Kellie, Th. Lamparter, J.A. MacKenzie, J.C. McGeorge, G.J. Miller, R.O. 
Owens, M. Sauer, R. Schneider and K. Spaeth, Phys. Lett. B 380 (1996) 
247;\newline
Th. Lamparter, J. Ahrens, J.R.M. Annand, I. Anthony, R. Beck, D. Branford, G.E. 
Cross, T. Davinson, P. Grabmayr, S.J. Hall, P.D. Harty, T. Hehl, J.D. Kellie, 
I.J.D. MacGregor, J.A. MacKenzie, J.C. McGeorge, G.J. Miller, R.O. Owens, M. 
Sauer, R. Schneider, K. Spaeth and G.J. Wagner, Z. Phys.  A 355 (1996) 1. 
\vspace{-3mm} 

\item \label{Mainz98}
I.J.D. MacGregor, T. T-H. Yau, J. Ahrens, J.R.M. Annand, R. Beck, D. 
Branford, P. Grabmayr, S.J. Hall, P.D. Harty, T. Hehl, J.D. Kellie, Th. 
Lamparter, M. Liang, J.A. MacKenzie, S. McAllister, J.C. McGeorge, R.O. Owens, 
J. Ryckebusch, M. Sauer, R. Schneider and D.P. Watts, Phys. Rev. Lett. 80 
(1998) 245. 
\vspace{-3mm} 

\item \label{Lund}
L. Isaksson, J-O. Adler, B-E. Andersson, K.I. Blomqvist, A. Sandell, B. 
Schr{\"o}der, P. Grabmayr, S. Klein, G. Mauser, A. Mondry, J.R.M. Annand, G.I. 
Crawford, J.C. McGeorge and  G.J. Miller, Proc. Second Workshop on 
Electromagnetically Induced Two-Nucleon Emission, ed. by J. Ryckebusch and M. 
Waroquier, Gent (1995) p. 301;\newline
L. Isaksson, Ph.D. thesis,  University of Lund, (1996).
\vspace{-3mm} 

\item \label{Oset}
R.C. Carrasco and E. Oset, Nucl. Phys.  A 536 (1992) 445;\newline
R.C. Carrasco, E. Oset and L.L. Salcedo, Nucl. Phys.  A 54 (1992) 585;\newline
R.C. Carrasco, M.J. Vicente Vacas and E. Oset, Nucl. Phys.  A 570 (1994) 701.
\vspace{-3mm} 

\item \label{Mainz1}
J. Ahrens, J.R.M. Annand, R. Beck, D. Branford, P. Grabmayr (spokesperson), 
S.J. Hall, T. Hehl, D.J. Ireland, J.D. Kellie, I.J.D. MacGregor, M. Mayer, 
J.C. McGeorge, F.A. Natter, R.O. Owens, M. Sauer, G.J. Wagner and S. 
Wunderlich, MAMI proposal Nr: A2/4-97.
\vspace{-3mm} 

\item \label{WAGP}
P. Wilhelm, H. Arenh{\"o}vel, C. Giusti, and F.~D.~Pacati, Z. Phys.  A 359 
(1997) 467.
\vspace{-3mm}

\item \label{LEGS}
R. Lindgren, V. Gladyshev, H. Baghaei, A. Caracappa, A. Cichocki, R. Finlay, 
T. Gresko, K. Hicks, S. Hoblit, M. Khandakar, O. Kistner, M. Lucas, L. Miceli, 
B. Norum, J. Rappaport, A. Sandorfi, R. Sealock, L. Smith, C. Thorn, S. 
Thornton. C. Whisnant, C. Giusti, F.D. Pacati and J. Ryckebusch, to be 
published 
\vspace{-3mm}

\item \label{MS93a}
H.M{\"u}ther and P.U. Sauer, Computational Nuclear Physics,
eds. K.-H. Langanke, J.A. Maruhn and S.E. Koonin, (Springer, New York, 1993).
\vspace{-3mm}

\item \label{Peccei}
R.D. Peccei, Phys. Rev. 176 (1968) 1812; 181 (1969) 1902.
\vspace{-3mm}

\item \label{WWA}
Th. Wilbois, P. Wilhelm and H. Arenh{\"o}vel, Phys. Rev.  C 54 (1996) 3311.
\vspace{-3mm}

\item \label{BM}
B.H. Bransden and R.G. Moorhouse, The Pion-Nucleon System, 
(Princeton, University Press, Princeton, 1973).
\vspace{-3mm}
     
\item \label{WNA}
P. Wilhelm, J.A. Niskanen and H. Arenh{\"o}vel, Nucl. Phys. A 597 (1996) 613.
\vspace{-3mm}

\item \label{C14}
F. Ajzenberg-Selove, Nucl. Phys.  A 523 (1991) 1.
\vspace{-3mm}

\item \label{Nad}
A. Nadasen, P. Schwandt, P.P. Singh, W.W. Jacobs, A.D. Bacher, P.T. Debevec,
M.D. Kaitchuk and J.T. Meek, Phys. Rev.  C 23 (1981) 1023.
\vspace{-3mm}

\item \label{ES}
L.R.B. Elton and A. Swift, Nucl. Phys.  A 94 (1967) 52.
\vspace{-3mm}

\item \label{GD}
C.C. Gearhart, Ph.D. thesis,  Washington University, St. Louis (1994);
C.C. Gearhart and W.H. Dickhoff, private communication.
\vspace{-3mm}
                        
\item \label {Clark}
J.W. Clark, {\em in\/} The Many-body problem: Jastrow correlations versus
Brueckner theory, ed. by R. Guardiola and J. Ros, Lecture Notes in Physics,
vol. 138 (Springer, Berlin 1981) p. 184.
\vspace{-3mm}

\item \label{OMY}
T. Ohmura, N. Morita and M. Yamada, Prog. Theor. Phys.  15 (1956) 222.
\vspace{-3mm}

\end{list}

\newpage
\begin{table}
\begin{tabular}{|cccccccc|}
\hline
& $J^\pi \,\, T$ & hole   &   & $n$ & $N$ & &\\
\hline
& 1$^+_1$ 0 & ($p_{1/2}$)$^{-2}$ & $^3S_1$; L = 0 & 1 & 0 & 0.14 &\\
&                     & &  $^3S_1$; L = 0 & 0 & 1 & $-$0.14 &\\
&                     & &  $^3S_1$; L = 2 & 0 & 0 & 0.61 &\\
&                     & &  $^1P_1$; L = 1 & 0 & 0 & $-$0.47 &\\
&                     & &  $^3D_1$; L = 0 & 0 & 0 & $-$0.61 &\\
\hline
&0$^+$ 1  & ($p_{1/2}$)$^{-2}$&$^1S_0$; L = 0 & 1 & 0 & $-$0.41 &\\
&                     & & $^1S_0$; L = 0 & 0 & 1 & 0.41 &\\
&                     & & $^3P_1$; L = 1 & 0 & 0 & 0.82 &\\
\hline
&1$^+_2$ 0  & ($p_{1/2}p_{3/2}$)$^{-1}$&$^3S_1$; L = 0 & 1 & 0 & $-$0.54 &\\
&                     & & $^3S_1$; L = 0 & 0 & 1 & 0.54 &\\
&                     & & $^3S_1$; L = 2 & 0 & 0 & 0.30 &\\
&                     & & $^1P_1$; L = 1 & 0 & 0 & 0.47 &\\
&                     & & $^3D_1$; L = 0 & 0 & 0 & $-$0.30 &\\
\hline
&2$^+$ 0  & ($p_{1/2}p_{3/2}$)$^{-1}$&$^3S_1$; L = 2 & 0 & 0 & $-$0.71 &\\
&                     & & $^3D_2$; L = 0 & 0 & 0 & $-$0.71 &\\
\hline
&3$^+$ 0  & ($p_{3/2}$)$^{-2}$&$^3S_1$; L = 2 & 0 & 0 & 0.71 &\\
&                     & & $^3D_3$; L = 0 & 0 & 0 & $-$0.71 &\\
\hline
\end{tabular}

\bigskip
 
\caption[Table I]{
Proton-neutron removal amplitudes from $^{16}$O for
states of $^{14}$N in a h.o. approximation and for different relative
$^{2S+1}l_j$ states. $L$ is the CM angular momentum.
\label{tab:pn}
}
\end{table}
\begin{table}
\begin{tabular}{|ccccccccc|}
\hline
& $J^\pi \,\, T$ & hole  &   & $n$ & $N$ &   & $c^{\,\mathrm{i}}$&\\
\hline
& 0$^+$ 1 & ($p_{1/2}$)$^{-2}$ & $^1S_0$; L = 0 & 1 & 0 & $-$0.41 & $-$0.42 &\\
&                     & & $^1S_0$; L = 0 & 0 & 1 & 0.41 & 0.42 &\\
&                     & & $^3P_1$; L = 1 & 0 & 0 & 0.82 & 0.51&\\
\hline
&2$^+$ 1 & ($p_{1/2}p_{3/2}$)$^{-1}$&$^1S_0$; L = 2 & 0 & 0 & $-$0.58 & 
$-$0.49 &\\
&                     & & $^3P_1$; L = 1 & 0 & 0 & 0.29 & 0.18 &\\
&                     & & $^3P_2$; L = 1 & 0 & 0 & 0.50 & 0.31 &\\
&                     & & $^1D_2$; L = 2 & 0 & 0 & 0.58 & 0.49 &\\
\hline
&1$^+$ 1  & ($p_{1/2}p_{3/2}$)$^{-1}$&$^3P_0$; L = 1 & 0 & 0 & 0.58 & 0.44 &\\
&                     & & $^3P_1$; L = 1 & 0 & 0 & 0.50 & 0.38 &\\
&                     & & $^3P_2$; L = 1 & 0 & 0 & $-$0.65 & $-$0.50 &\\
\hline
\end{tabular}

\bigskip
 
\caption[Table II]{
Two-proton removal amplitudes from $^{16}$O for
states of $^{14}$C in a h.o. approximation and for different relative
$^{2S+1}l_j$ states. $L$ is the CM angular momentum. The amplitudes are 
compared with the corresponding coefficients $c^{\mathrm{i}}$ 
in Eq.~(\ref{eq:ppover}), obtained from the calculation of the spectral 
function of ref.~[\ref{Geurts}].
\label{tab:pp}
}
\end{table}

\begin{figure}
\caption[]{The differential cross section  of the reaction 
$^{16}\mathrm{O}(\gamma,\mathrm{pn})^{14}\mathrm{N}$, as a function of the 
angle $\gamma_{1}$, in coplanar symmetrical kinematics at $E_\gamma=100$ MeV, 
for the transitions to the low-lying states in $^{14}$N: $1^+_1$ 
($E_{2\rm{m}} = 22.96$ MeV), $0^+$ ($E_{2\rm{m}} = 25.27$ MeV), $1^+_2$ 
($E_{2\rm{m}} = 26.91$ MeV), $2^+$ ($E_{2\rm{m}} = 29.99$ MeV), $3^+$ 
($E_{2\rm{m}} = 34.01$ MeV). The optical potential is taken from 
ref.~[\ref{Nad}], the single-particle wave functions from ref.~[\ref{ES}] and 
the correlation function from ref.~[\ref{GD}]. Separate contributions of the 
one-body current (1-B) and of its sum with the two-body seagull (SEA) and 
pion-in-flight ($\pi$) currents are drawn. The solid lines give the total 
cross section, where also the $\Delta$ current ($\Delta$) is added. 
\label{fig:gnn1}
}
\end{figure}

\begin{figure}
\caption[]{The differential cross section  of the reaction 
$^{16}\mathrm{O}(\gamma,\mathrm{pn})^{14}\mathrm{N}$, as a function of the 
angle $\gamma_{1}$, in the same kinematics and for the same transitions as in 
Fig.~1. Optical potential, single-particle wave functions and correlation 
function as in Fig.~1. The solid lines are the same as in Fig.~1. Separate 
contributions of the $\Delta$, seagull, pion-in-flight currents and of the sum 
of seagull and pion-in-flight currents are drawn. 
\label{fig:gnn2}
}
\end{figure}

\begin{figure}
\caption[]{The differential cross section  of the reaction 
$^{16}\mathrm{O}(\gamma,\mathrm{pn})^{14}\mathrm{N}$, as a function of the 
angle $\gamma_{1}$, in coplanar symmetrical kinematics at $E_\gamma=300$ MeV, 
for the same transitions, under the same conditions and with same line 
convention as in Fig.~1. 
\label{fig:gnn3}
}
\end{figure}

\begin{figure}
\caption[]{The differential cross section  of the reaction 
$^{16}\mathrm{O}(\gamma,\mathrm{pn})^{14}\mathrm{N}$, as a function of the 
angle $\gamma_{1}$, in coplanar symmetrical kinematics at $E_\gamma=300$ MeV, 
for the same transitions, under the same conditions and with same line 
convention as in Fig.~2. 
\label{fig:gnn4}
}
\end{figure}

\begin{figure}
\caption[]{The differential cross section of the reaction 
$^{16}\mathrm{O}(\gamma,\mathrm{pn})^{14}\mathrm{N}$, as a function of the 
photon energy, for a fixed value of the angle $\gamma_{1}$, in coplanar 
symmetrical kinematics, for the same transitions as in Fig.~1. Optical 
potential, single-particle wave functions and correlation function as in 
Fig.~1. Line convention as in Fig.~1.
\label{fig:gnn5}
}
\end{figure}

\begin{figure}
\caption[]{The differential cross section of the reaction 
$^{16}\mathrm{O}(\gamma,\mathrm{pn})^{14}\mathrm{N}$, as a function of the 
photon energy, for a fixed value of the angle $\gamma_{1}$, in coplanar 
symmetrical kinematics,  for the same transitions, under the same conditions 
and with same line convention as in Fig.~2. 
\label{fig:gnn6}
}
\end{figure}

\begin{figure}
\caption[]{The photon asymmetry of the reaction 
$^{16}\mathrm{O}(\vec{\gamma},\mathrm{pn})^{14}\mathrm{N}$, as a function of 
the angle $\gamma_{1}$, in the coplanar symmetrical kinematics at 
$E_\gamma=100$ MeV, for the same transitions, under the same conditions and 
with same line convention as in Fig.~1. 
\label{fig:gnn7}
}

\end{figure}
\begin{figure}
\caption[]{The photon asymmetry of the reaction 
$^{16}\mathrm{O}(\vec{\gamma},\mathrm{pn})^{14}\mathrm{N}$, as a function of 
the angle $\gamma_{1}$, in the coplanar symmetrical kinematics at 
$E_\gamma=100$ MeV, for the same transitions, under the same conditions and 
with same line convention as in Fig.~2. 
\label{fig:gnn8}
}
\end{figure}

\begin{figure}
\caption[]{The photon asymmetry of the reaction 
$^{16}\mathrm{O}(\vec{\gamma},\mathrm{pn})^{14}\mathrm{N}$, as a function of 
the angle $\gamma_{1}$, in the coplanar symmetrical kinematics at 
$E_\gamma=300$ MeV, for the same transitions, under the same conditions and 
with same line convention as in Fig.~3. 
\label{fig:gnn9}
}
\end{figure}

\begin{figure}
\caption[]{The photon asymmetry of the reaction 
$^{16}\mathrm{O}(\vec{\gamma},\mathrm{pn})^{14}\mathrm{N}$, as a function of 
the angle $\gamma_{1}$, in the coplanar symmetrical kinematics at 
$E_\gamma=300$ MeV, for the same transitions, under the same conditions and 
with same line convention as in Fig.~4. 
\label{fig:gnn10}
}
\end{figure}

\begin{figure}
\caption[]{The differential cross section  of the reaction 
$^{16}\mathrm{O}(\gamma,\mathrm{pp})^{14}\mathrm{C}$, as a function of the 
angle $\gamma_{1}$, in coplanar symmetrical kinematics at $E_\gamma=100$ MeV, 
for the transitions to the low-lying states in $^{14}$C: $0^+_1$ ($E_{2\rm{m}} 
= 22.33$ MeV), $2^+$ ($E_{2\rm{m}} = 30.00$ MeV), $1^+$ ($E_{2\rm{m}} = 33.64$ 
MeV). The defect functions for the Bonn-A potential and the optical potential 
of ref.~[\ref{Nad}] are used. Separate contributions of the one-body and 
two-body $\Delta$ currents are drawn. The solid lines give the 
total cross sections, produced by the sum of the one-body and $\Delta$ 
currents. 
\label{fig:gnn11}
}
\end{figure}

\begin{figure}
\caption[]{The differential cross section  of the reaction 
$^{16}\mathrm{O}(\gamma,\mathrm{pp})^{14}\mathrm{C}$, as a function of the 
angle $\gamma_{1}$, in the same kinematics and for the same transitions as in 
Fig.~11. Defect functions and optical potential as in Fig.~11. The solid lines 
are the same as in Fig.~11. Separate contributions of different partial waves 
of relative motion are drawn. 
\label{fig:gnn12}
}
\end{figure}

\begin{figure}
\caption[]{The differential cross section  of the reaction 
$^{16}\mathrm{O}(\gamma,\mathrm{pp})^{14}\mathrm{C}$, as a function of the 
angle $\gamma_{1}$, in coplanar symmetrical kinematics at $E_\gamma=300$ MeV, 
for the same transitions, under the same conditions and with same line 
convention as in Fig.~12. 
\label{fig:gnn13}
}
\end{figure}

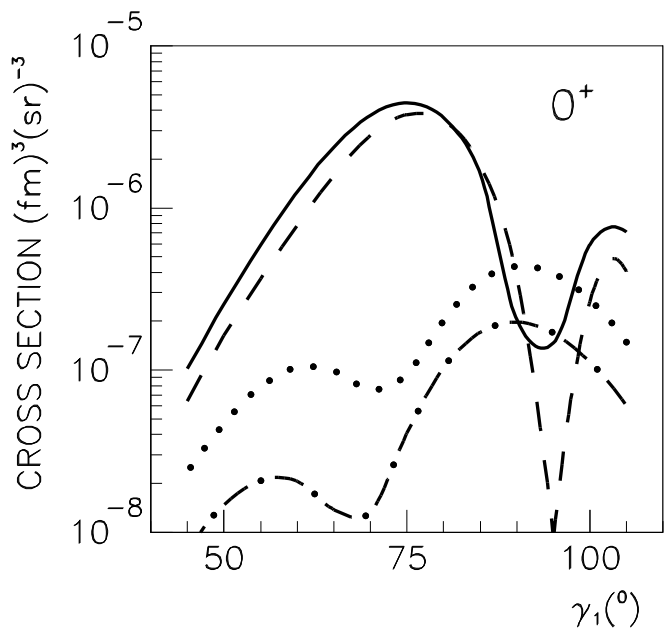
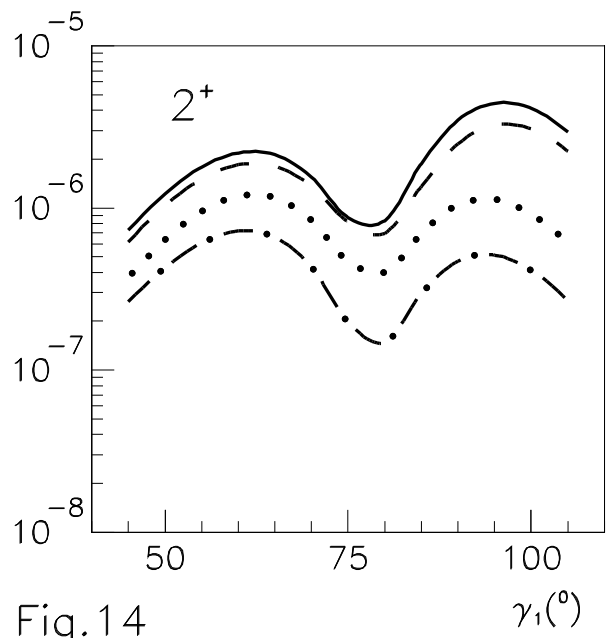
\begin{figure}
\caption[]{The differential cross section  of the reaction 
$^{16}\mathrm{O}(\gamma,\mathrm{pp})^{14}\mathrm{C}$, as a function of the 
angle $\gamma_{1}$, in coplanar symmetrical kinematics at $E_\gamma=100$ MeV, 
for the transitions to the $0^+$ and $2^+$ states in $^{14}$C. The solid 
and dashed lines are calculated with the defect functions of the Bonn-A and 
Reid potentials, respectively; the dotted and dot-dashed lines give the 
corresponding separate contributions of the one-body current. The optical 
potential is taken from ref.~[\ref{Nad}]
\label{fig:gnn14}
}
\end{figure}

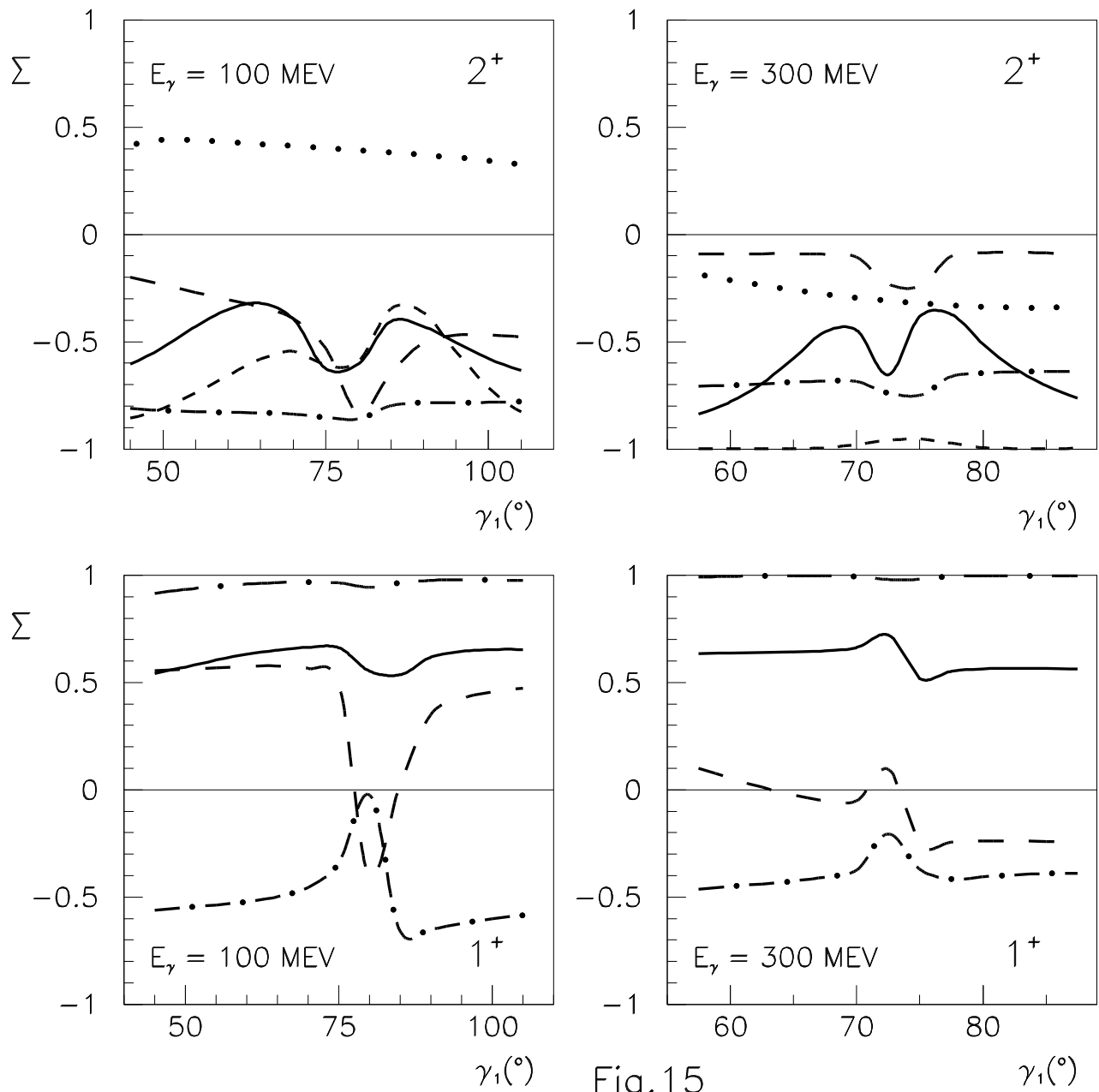
\begin{figure}
\caption[]{The photon asymmetry of the reaction 
$^{16}\mathrm{O}(\vec{\gamma},\mathrm{pp})^{14}\mathrm{C}$, as a function of 
the angle $\gamma_{1}$, for the transitions the $2^+$ and $1^+$ states in 
$^{14}$C in the same kinematics as in Fig.~12. Defect functions, optical 
potential and line convention as in Fig.~12. 
\label{fig:gnn15}
}
\end{figure}

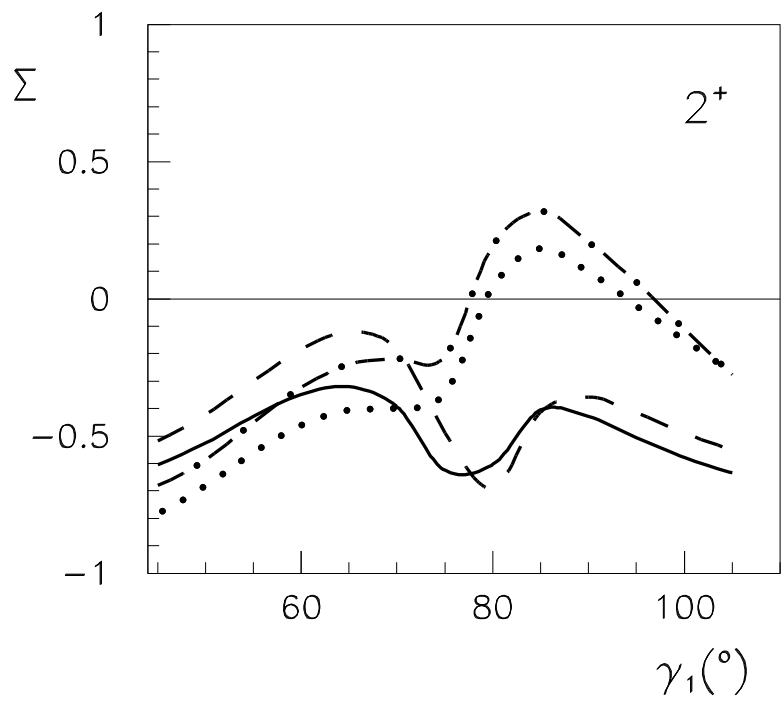
\begin{figure}
\caption[]{The photon asymmetry of the reaction 
$^{16}\mathrm{O}(\vec{\gamma},\mathrm{pp})^{14}\mathrm{C}$, as a function of 
the angle $\gamma_{1}$, in coplanar symmetrical kinematics at $E_\gamma=100$ 
MeV, for the transition to the $2^+$ state in $^{14}$C. Optical potential and 
line convention as in Fig.~14. 
\label{fig:gnn16}
}
\end{figure}

\end{document}